\documentclass[lettersize,journal]{IEEEtran}
\usepackage{amsmath,amsfonts}
\usepackage{algorithmic}
\usepackage{algorithm}
\usepackage{array}
\usepackage[caption=false,font=normalsize,labelfont=sf,textfont=sf]{subfig}
\usepackage{textcomp}
\usepackage{stfloats}
\usepackage{url}
\usepackage{verbatim}
\usepackage{graphicx}
\usepackage{cite}
\usepackage{xcolor}
\usepackage[normalem]{ulem}
\begin{document}

\title{Measuring dark currents in multiple cryogenic SiPMs with sub-pA sensitivity using an automated IV multiplexer}

\author{Lucas Darroch$^{1}$, Eamon Egan$^{1}$, Marc-Antoine Leclerc$^{1}$, 
Thomas McElroy$^{1}$, and Thomas Brunner$^{1}$%
\thanks{$^{1}$Department of Physics, McGill University, Montreal, QC H3A 2T8, Canada.}%
\thanks{Lucas Darroch's current affiliation: is with the Wright Laboratory, Yale University, New Haven, USA (e-mail: lucas.darroch@yale.edu).}%
\thanks{Marc-Antoine Leclerc's current affiliation: Medical Physics Unit, McGill University, Montreal, Canada.}%
\thanks{Thomas McElroy's current affiliation: PulseMedica, Edmonton, Canada.}
}


\maketitle

\begin{abstract}
We present the design of an automated current-voltage (IV) multiplexer (MUX) that enables accurate measurement of the dark current in cryogenic silicon photomultipliers (SiPMs), achieving a sensitivity equivalent to detecting less than one avalanche per second. Dynamic pulse-counting measurements were used as a benchmark for reconstructing the dark current in static IV measurements. The IV-MUX features 15 channels on a single board and up to seven boards can be connected in parallel under the control of one Arduino microcontroller. To minimize leakage and enhance performance, the layout includes guard rings and high-isolation relays, enabling resolution of currents as small as 49 fA. The IV-MUX can be integrated into systems designed for IV or pulse-counting measurements, enabling seamless switching between IV and pulse-counting modes. Moreover, the IV-MUX is vacuum-compatible, validated by testing an SiPM array in a cryostat. This feature reduces the need for multiple feedthroughs when testing sensor arrays in vacuum. The design is open source and can be used to facilitate rapid and automated testing of SiPMs or similar low-current devices in one measurement cycle.
\end{abstract}


\section{Introduction}
\label{sec:1introduction}

\IEEEPARstart{S}{ilicon} photomultipliers (SiPMs) have emerged as a popular alternative to traditional photodetectors, such as photomultiplier tubes or photodiodes. An SiPM is an array of single-photon avalanche diodes (SPADs), each operated in reverse bias mode above its breakdown voltage. SiPMs have high photo-detection efficiency, fast response, high gain, low noise, and excellent single photon resolution \cite{ref:renker2009advances,ref:ghassemi2017mppc,ref:acerbi2019understanding}. These features make SiPMs useful for counting the number of photons in a scintillation signal by dynamically measuring the SiPM response. Several next-generation particle physics experiments are implementing large arrays of SiPMs, totaling several square-meters in size, immersed in noble liquids as the photosensors of choice \cite{ref:kharusi2018nexo,ref:aalseth2018darkside,ref:dune2020deep}. This necessitates accurate methods to efficiently test thousands of individual SiPMs. 

Current-voltage (IV) measurements are commonly used to characterize the electrical properties of semiconductors, such as diodes, transistors, and solar cells. The process involves applying a voltage across the device and measuring the resulting current. In SiPMs, IV characteristics are valuable for verifying compliance with specification requirements. The IV response of SiPMs can alternatively be reconstructed through dynamic measurements \cite{ref:acerbi2019understanding}, using models that predict how these measurements vary with voltage \cite{ref:otte2017characterization, gallina2019characterization, gallina2021development}. Hence, the IV response of an SiPM encodes information about the optical, electrical, and nuisance parameters that describe its dynamic response. Measuring the IV response may be the most practical method for bulk SiPM characterization in experiments that apply them at large scale.  

The photodetection efficiency (PDE), single photoelectron (SPE) charge, correlated avalanche (CA) noise, and dark count rate (DCR) all contribute to determining the incident light flux on an SiPM. In SiPM technologies designed for noble liquid detectors, the PDE, SPE charge, and CA noise exhibit a weak temperature dependence \cite{ref:gallina2019characterization, ref:wang2021characterization, ref:alvarez2024measurement, ref:borden2024characterization}, whereas the dark count rate (DCR) strongly depends on temperature \cite{ref:piemonte2019overview}. Although there is a general understanding of how temperature affects these parameters, cryogenic testing is often necessary to sufficiently characterize SiPMs for cryogenic applications, such as noble liquid detectors. Additionally, there is a random probability for SPADs to contain structural or contamination-related defects that cause a high DCR \cite{cova1996avalanche, giudice2007high, parent2024wafer}. Moreover, the location of defective SPADs can be spatially correlated, which can result in clusters of defects within a device \cite{engelmann2018spatially, mclaughlin2021spatially}. At moderate temperatures, identifying SiPMs with defective SPADs through IV measurements can be challenging because the summed DCR from normally functioning SPADs obscures the contribution from defective SPADs. However, at cryogenic temperature, the DCR contribution from defective SPADs may remain disproportionately high, as structural or contamination-related defects are less affected by cooling. Identifying and rejecting defective or poorly operating devices prior to detector assembly is crucial to ensure that the dark noise contribution remains below design specifications.

Cryogenic testing of SiPMs is more efficient when large batches are processed in a single cooldown phase, mostly due to the time required to cycle the cryostat. The dark current in an SiPM at cryogenic temperature is very small and requires a sensitive ammeter and specialized coaxial/triaxial vacuum feedthroughs. The high cost per channel for measuring the dark current in cryogenic SiPMs motivated the development of a vacuum-compatible IV multiplexer (IV-MUX) capable of routing the signals for up to 105 SiPMs through a single triaxial vacuum feedthrough to an ammeter with minimal leakage current. This is particularly challenging, as recent work by researchers at FBK \cite{ref:acerbi2023nuv} demonstrates that measuring low currents at cryogenic temperatures is difficult; their resolution, on the order of 1 pA, limits their ability to detect dark current above the noise from leakage currents when testing a single SiPM. In contrast, our IV-MUX achieves sub-pA resolution while multiplexing multiple SiPMs, all while managing the added complexity of measuring low currents across multiple devices with vacuum compatible materials. We describe the design of the IV-MUX and present results from two sets of performance tests: (1) measurements in vacuum to validate multi-channel operation with a single feedthrough, and (2) precision measurements in liquid nitrogen (LN) to assess sub-pA sensitivity. In the latter case, dynamic pulse-counting measurements are compared to IV measurements to benchmark low-current SiPM behavior and validate the IV-MUX in regimes where conventional systems lack sufficient resolution. Together, these results demonstrate that the IV-MUX is a reliable and efficient tool for high-throughput characterization of cryogenic SiPMs.

\section{Design}\label{sec:2design}

To facilitate high-throughput IV testing of cryogenic SiPMs, our requirements are: (1) switching bias voltage for up to 105 SiPMs to provide scalability for large systems; (2) ensuring that the signal-noise remains below 1 pA to maintain measurement accuracy; (3) minimizing the number of vacuum feedthroughs to reduce complexity and leakage risks in cryogenic environments; (4) the ability to measure current using either the bias node or a dedicated current-sense node for flexibility in measurement configuration; and (5) minimizing operator intervention after the initial setup to ensure efficient operation in batch testing scenarios. The system's capacity for 105 SiPMs was chosen to balance practical needs for testing large SiPM arrays with the simplicity of the design. 

\begin{figure}[t]
    \centering
    \includegraphics[width=\columnwidth]{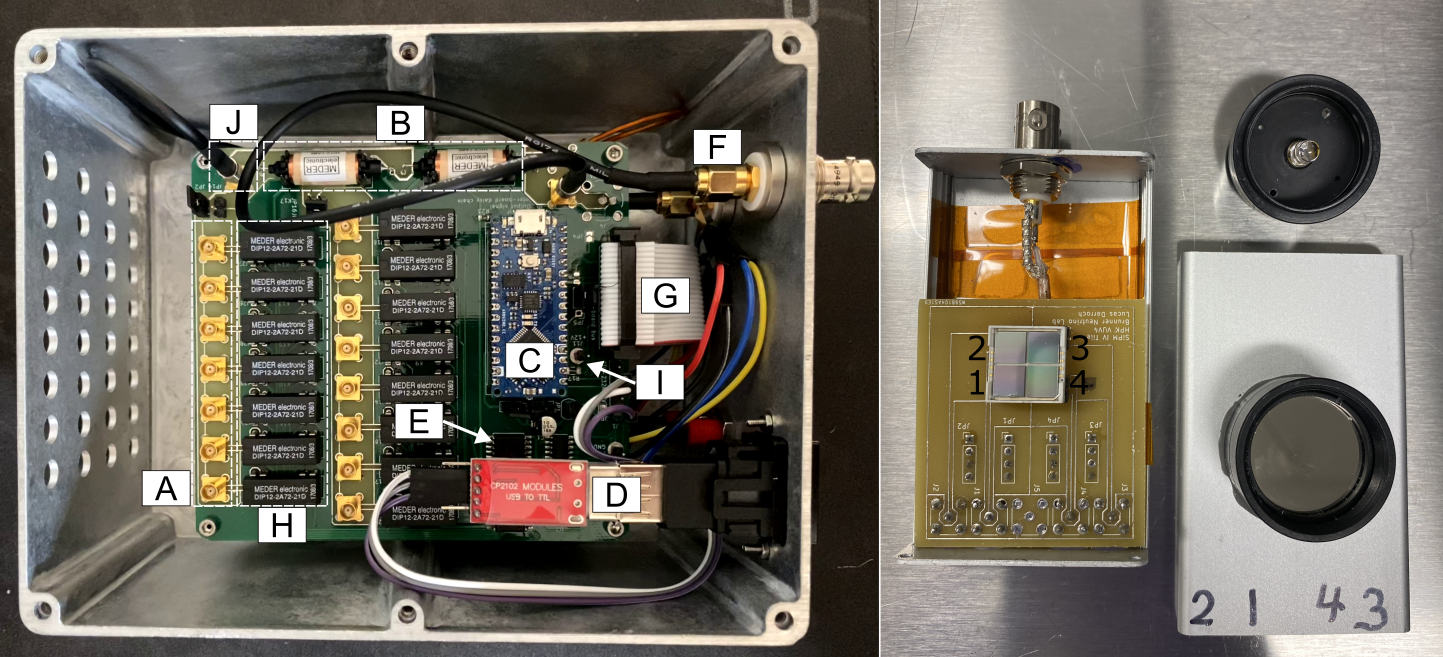}
    \caption{Left: Annotated IV-MUX in shielded enclosure for bench-top use. (A) 15 bias connectors, (B) sense line relays, (C) Arduino, (D) serial communication, (E) shift registers, (F) triaxial readout, (G) board interconnect, (H) bias line relays, (I) 12 V board power and bias power, (J) sense line input for mode B (described in text). Right: A Hamamatsu SiPM inside an aluminum enclosure, mounted on a custom PCB designed for routing signals in either mode A or mode B of the IV-MUX, which is described in Section \ref{sec:2design}. Each quadrant of the device is segmented, allowing four independent channels, which are labeled. The top of the enclosure, which includes an optical port and a demountable LED, is shown next to the SiPM. The SiPM is used for measurements in LN, which are presented in Section \ref{sec:4airtest}.}
    \label{fig:airmux}
\end{figure}

To meet these criteria, we designed an Arduino-controlled high-isolation IV-MUX, which is shown in Figure \ref{fig:airmux}, on the left. It serves as a cost-effective way to improve the efficiency of an existing IV measurement system through multiplexing. Each IV-MUX board supports 15 channels and the system can be expanded to 105 channels by interconnecting up to seven boards, all controlled as a single unit by one Arduino Nano Every module. To accommodate even higher channel counts, multiple seven-board (105 channel) sets, each controlled by one Arduino, can be connected in parallel, or the control bus can be expanded to handle more boards. 

Each channel is switched to a shared electrometer input via a single measurement path, minimizing inter-channel variation due to readout circuitry. To suppress leakage and maintain sub-pA resolution across all channels, the design employs guarded traces to reduce parasitic coupling and surface leakage, along with separate sense and bias cables for each device. Relay selection prioritized low leakage current and high isolation to preserve signal integrity in a vacuum environment. Critical aspects of the design are discussed in the following sections.

\subsection{Minimizing leakage current}

Achieving sub-pA resolution in dark current measurements requires careful mitigation of leakage and noise sources. The dominant contribution typically arises from electromagnetic interference, which is addressed through shielding, separation of high- and low-side signal paths, and the use of triaxial cables. Additional leakage could arise from surface contamination on the PCB, such as solder flux, oils, or particulates, that can form weakly conductive paths between traces. Standard cleaning techniques for vacuum electronics are applied to minimize these effects, and vacuum standards were adhered to in the handling of any electronics components after their cleaning. Relay open-contact resistance and leakage from power rails into the sense node can also contribute current at sub-pA levels. To suppress these effects, the design incorporates a guard node, dedicated sense and bias lines, and relay control strategies to isolate deactivated channels. These features are described in the following section.

Leakage current is intercepted using a circuit node called a guard node. The guard node is kept at a voltage within a few mV of the potential of sensitive traces, which is intended to separate nodes carrying relatively high voltages from nodes that are sensitive to leakage current. In this design, the guard node is close to ground potential. PCB traces that are connected to the guard node and free of solder mask form a protective ring around sensitive pads and traces. This structure, known as a guard ring, serves to shield the sensitive components from external influences and minimize undesirable effects such as leakage currents or electromagnetic interference \cite{ref:kester1992amplifierappguide}.

Each bias channel is controlled by a double pole single throw (DPST) reed relay (Standex Meder Electronics, DIP12-2A72-21D), where the two poles are arranged in series. Each bias voltage connection is surrounded by a guard ring to prevent current from leaking into a deactivated bias channel from nearby active channels or from the 12 V relay coil pins. The guard ring extends from the point where the bias voltage emerges from the bias voltage switching relay to where it connects to the MCX connector center pin. Likewise, a guard ring protects the current-sense node to minimize potential interference. The circuit node between the two relay poles is connected to the guard node with a 100 k$\Omega$ resistor, and intercepts current which may leak through the first relay contact when the relay is off. 

\begin{figure*}[t]
\centering
\includegraphics[width=\textwidth]{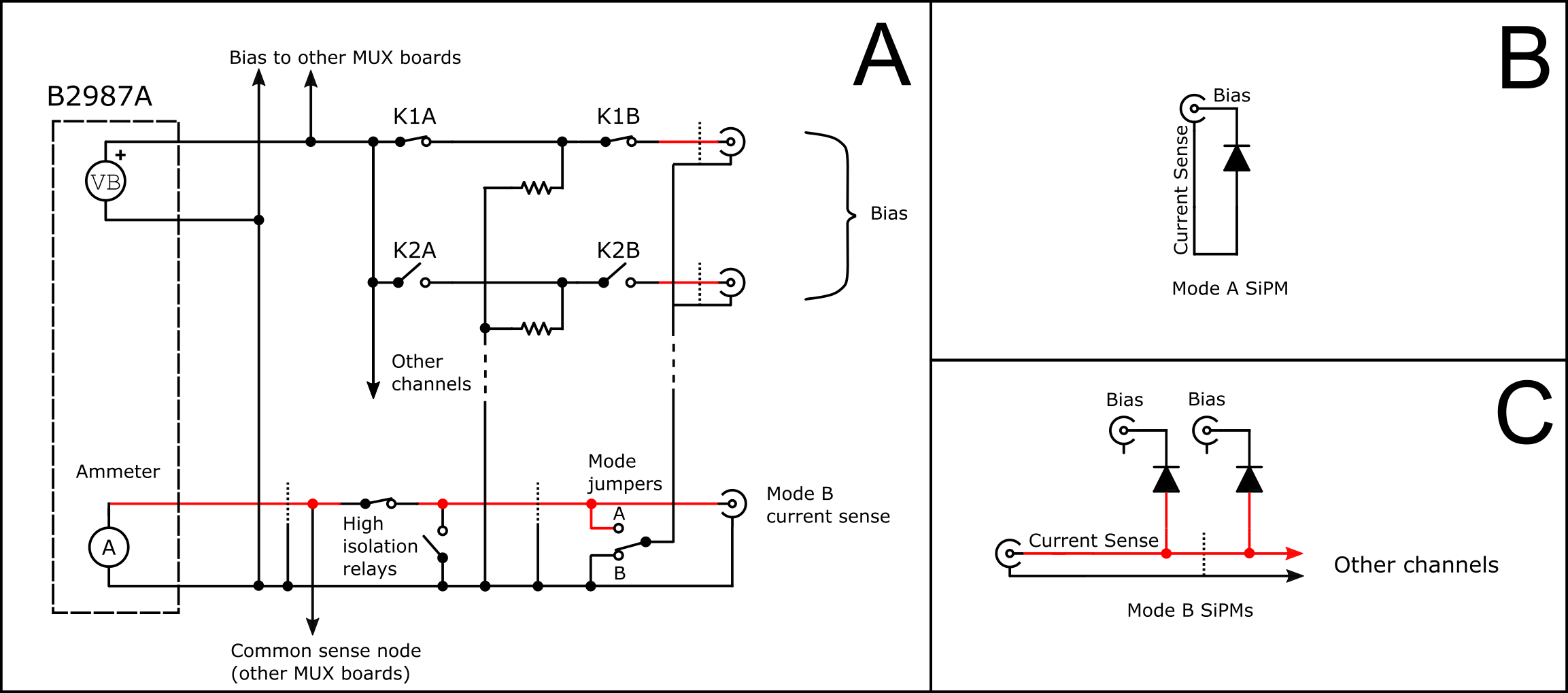}
\caption{(A) A simplified circuit diagram showing the IV-MUX switches and current paths on the MUX. Red traces are guarded by the nodes indicated with the dotted lines. A combined sense node configuration is shown for use with a single ammeter - isolating this interconnect would allow each board to be used with a separate ammeter. (B) A single SiPM connected in mode A. (C) Two SiPMs connected in mode B.}
\label{fig:muxschem}
\end{figure*}

To support multiplexing between boards, the current-sense node uses two high-isolation single-pole reed relays (Standex Meder Electronics, HI12-1A85): one shunting the current-sensing trace to ground, the other interrupting the connection between the current-sensing trace and the connection to the electrometer and to the corresponding outputs of any other board in the system. The current-sense node is the most leakage-sensitive part of the circuit, and a high-isolation relay is used primarily to prevent leakage from the 12 V relay activation pins. In a multi-board setup, when a given IV-MUX board is active, its series relay is closed and its shunt relay is open; when inactive, the states are reversed. Each board in a multi-board setup can be configured with either a common or isolated current-sense node, allowing compatibility with both single and multi-channel ammeters. A simplified circuit diagram of the IV-MUX is shown in Figure \ref{fig:muxschem}. The schematics and layout of the IV-MUX are available on a public GitHub repository \cite{ref:muxgit}.

\subsection{Control}

The IV-MUX system supports multi-board expansion via a shared SPI-style control bus, enabling a single Arduino Nano Every to manage up to seven boards. The firmware for the Arduino includes a feature to sleep the microcontroller when not in use to reduce power consumption. Each board includes addressable shift registers (two SN74HC595 per board) that control relay switching for 15 bias lines and one sense line, eliminating the need for additional microcontrollers. This architecture allows up to 105 SiPMs to be addressed automatically, reducing manual intervention and improving measurement efficiency. Boards are connected via a ribbon cable carrying control and power signals, while the sense path is handled separately via triaxial cables. Boards can be configured in a daisy-chain fashion, where multiple boards share a common sense output, or isolated to provide independent sense outputs for parallel readout with multi-channel ammeters. This modularity provides users the flexibility to balance measurement speed, cryostat feedthrough constraints, and instrumentation resources depending on their application.

The firmware for the Arduino is available in the public IV-MUX
repository, which includes supplemental documentation \cite{ref:muxgit}. We developed a Python control program that facilitates automated channel switching and measurement using the IV-MUX and the Keysight B2987A electrometer, which we use in our experiments. The accompanying Python control program automates the channel switching and measurement process, sending commands to both the IV-MUX and the electrometer. For a typical measurement, a TTL signal is sent to the IV-MUX to set a channel, then SCPI commands are sent to the electrometer to perform a series of measurements, and then the buffer-data is saved to a computer. This system ensures efficient data collection and reduces operator workload. 

\subsection{Device connectivity}

The IV-MUX can connect to SiPMs in two different ways, which we refer to as mode A and mode B. Mode A works with fully isolated SiPM devices. In this mode, the center and shield of the IV-MUX bias connector are connected directly to the SiPM cathode and anode, respectively, through a coaxial cable, and the coaxial shields of all the coaxial cables are connected to the current-sense node. Mode A was designed to be compatible with setups where the SiPMs are individually connected to readout cables, and do not share a common anode. In this configuration, low-current measurement is limited because the SiPM current is carried on the shield of the coaxial cable, and is therefore subject to electromagnetic interference. However, this effect can be mitigated by operating the IV-MUX (and the SiPM readout cables) inside of a shielded environment, such as a cryostat. Additionally, center to shield leakage current from the coaxial connectors and cables adds to the SiPM current signal, and the current-sensing trace may pick up noise current through electrostatic induction because it is unshielded.

Mode B works with interconnected devices that share a common anode. The center of the bias connector is connected to the SiPM cathode, but the shield is at ground potential. The common anode for all SiPMs in mode B is connected to the current-sense node through a dedicated coaxial cable. In this configuration, the IV-MUX can be operated in an unshielded environment because the current-sense node is shielded. A dedicated coaxial cable for current-sensing also eliminates leakage from the bias through the cable's dielectric. Additionally, using separate cables for bias supply and current-sensing allows integration of an amplifier coupled to a digitizer/oscilloscope on the high-side of the SiPM for dynamic measurements, de-coupled from the ammeter on the low-side, which increases the flexibility of the measurement system.

The IV-MUX was tested in two applications: first, both modes were tested with an SiPM array inside a cryostat. Second, the IV-MUX was tested with a bench-top setup using an SiPM in LN. Both IV and dynamic measurements are compared.

\section{IV-MUX performance in vacuum}\label{sec:3vactest}
\subsection{Hardware setup}

The IV-MUX was assembled with SAC305 solder and washable solder flux, then cleaned and sonicated in isopropanol. The IV-MUX was mounted onto a black SLA 3D-printed bracket that is screwed to the inside of a con-flat flange; black SLA resin does not fluoresce, which is important when testing photodetectors that are sensitive to single photons, such as SiPMs. The con-flat flange includes two feedthroughs: a 9-pin Subminiature C feedthrough for control and bias signals and a triaxial feedthrough for the current signal (Accu-Glass Products).

IV measurements of an SiPM array were performed at 165 K using both modes (A and B) of the IV-MUX. The SiPM array, a 16-channel Hamamatsu S13775-9121 mini-tile, was soldered to a jumper-configurable PCB designed for compatibility with modes A and B of the multiplexer. Similar to the IV-MUX, all current-sensing traces on the PCB were enclosed with a guard trace, and an additional connector was included for current-sensing in mode B. The PCB was mounted onto a copper chuck and positioned atop a custom cryostat cold-plate. The temperature of the SiPM array was monitored using calibrated platinum RTDs (Lakeshore Cryogenics, Pt100) that were installed within an aluminum lug and securely bolted to the front of the SiPM PCB. Apiezon N vacuum grease (Apiezon) was applied to the bottom of the aluminum lug to increase thermal conductivity between the lug and the PCB.

\begin{figure}[t]
    \centering
    \includegraphics[width=\columnwidth]{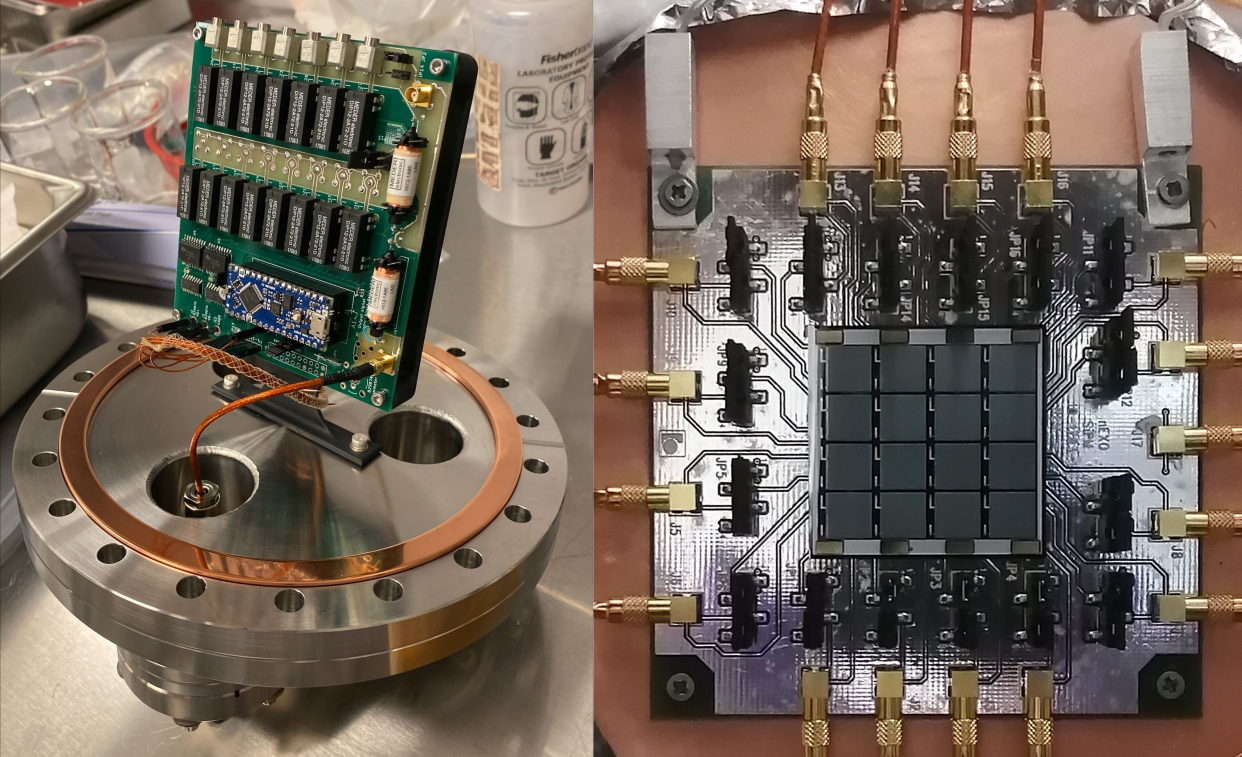}
    \caption{Left: The IV-MUX prepared for vacuum use. The system uses one triaxial connection to the electrometer for current readout, and one 9-pin Subminiature C feedthrough for power, bias voltage and communication with the Arduino. In this configuration, it is equipped for measurement of up to 15 SiPMs. Right: A Hamamatsu SiPM mini-tile mounted atop a custom PCB configured to interface with both modes of the multiplexer, as explained in Section \ref{sec:2design}. Header pin jumpers are used to switch between the two modes. The 16-channel SiPM array was used for measurements within a cryostat presented in this section. The two aluminum lugs housing RTDs are in the top-left and top-right corners of the PCB.}
    \label{fig:vacmux}
\end{figure}

The IV-MUX was placed in an adjacent chamber within the same vacuum setup as the cryostat, which was evacuated using a turbomolecular pump backed by a dry scroll pump. The temperature of the IV-MUX was measured at two on-board devices that we suspected could overheat in vacuum: a DPST relay, and the Atmega4809 microcontroller on the Arduino. The temperature of a DPST relay was measured by gluing a temperature sensor to its body, and the temperature of the ATmega4809 microcontroller was measured using its built-in sensor. All cables inside the setup were Kapton-insulated (Accu-Glass Products) and cleaned for vacuum. Inside the vacuum chamber, coaxial cables were used for biasing the SiPMs and conducting the current. Outside the vacuum chamber, a triaxial cable was used to carry the current to the electrometer (Keysight, B2987A), which was used to measure the current and supply bias voltage. A picture of the IV-MUX mounted on a con-flat flange (left) and the 16-channel SiPM array (right) are shown in Figure \ref{fig:vacmux}. 

\subsection{Measurements}

\begin{figure}[t]
    \centering
    \includegraphics[width=\columnwidth]{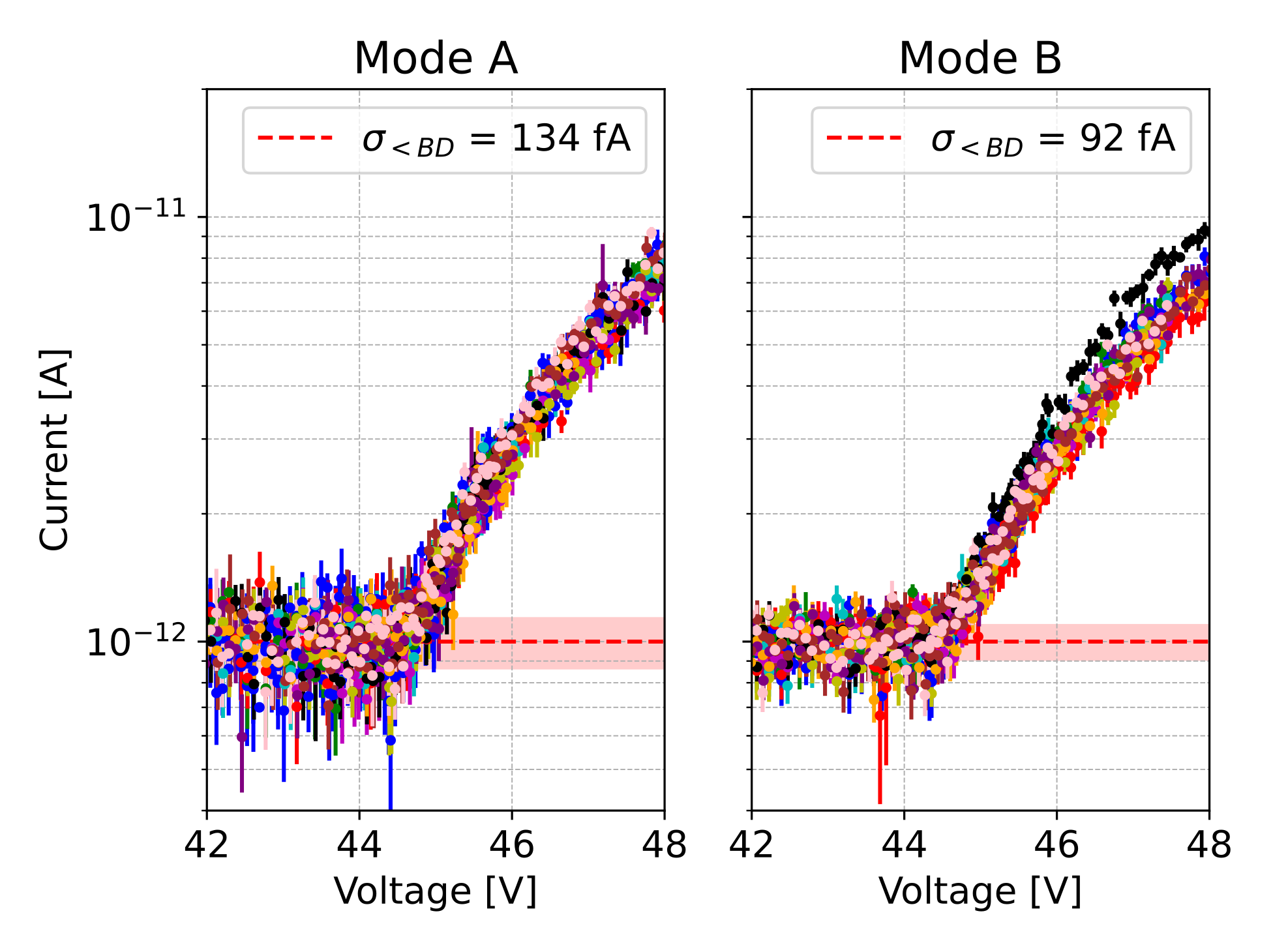}
    \caption{Dark IV measurements of 12 SiPMs using mode A (left) and mode B (right) of the IV-MUX. The standard deviation of the pre-breakdown region, which provides an estimate of the minimum signal resolution, was calculated and is shown in the legend. A constant value was added to each IV curve to align the pre-breakdown region for all SiPMs, which facilitates visual comparison. The uncertainties are a combination of statistical and instrumental uncertainties, added in quadrature.}
    \label{fig:muxsipmdata}
\end{figure}

The cryostat was evacuated to below $10^{-6}$ mbar, measured with a cold-cathode gauge (Pfeiffer Vacuum, PKR 360), and then cooled down until the RTD next to the SiPM array stabilized to 165 K. The temperature stability during each measurement was within 2 mK, ensuring consistent IV behavior and minimizing thermal effects on the SiPM’s response during the test. The DPST relay temperature stabilized at 315 K with a single relay active and at 350 K with all relays active, while the Arduino stabilized at 310 K and 330 K in each scenario, respectively. The SiPM and IV-MUX were configured first in mode A, then in mode B, and the IV curves were recorded for each channel in the dark, with the cold-cathode gauge powered off. Figure \ref{fig:muxsipmdata} shows the dark IV curves for 12 SiPMs in both modes. The remaining 4 SiPMs on the array were connected differently and not used in this analysis. Some IV data points in mode A were omitted due to the auto-ranging feature of the electrometer switching during the measurement, which caused a discontinuity in the IV curve due to a different zero/offset value. Data points following a range change were identified irrespective of the IV value through a time delay $>$ 1 second between measurements that is caused by a processing delay when entering a new measurement range (typical measurements were spaced by $\sim$ 50 ms within a fixed range). A fixed range was used in subsequent measurements to prevent this issue.

The pre-breakdown region of the IV curve is the SiPM linear-mode where each SPAD has a proportional response to the radiant flux \cite{ref:campbell2004recent}. The low-gain and DCR of a cryogenic SiPM make the change in linear-mode current as a function of voltage effectively zero, given the resolution of the present experimental setup. Therefore, a small region in linear-mode can be averaged to determine a minimum signal resolution for each mode. The pre-breakdown region between 42 and 44 V was averaged, and each IV curve was offset by adding a constant value such that the mean value of the pre-breakdown region was 1 pA (for visual comparison). The standard deviation of the data in the pre-breakdown region is $\sigma_{< \mathrm{BD}}$ = 134 fA for mode A, and $\sigma_{< \mathrm{BD}}$ = 92 fA for mode B, demonstrating improved performance with mode B.

\section{IV and Dynamic Response}\label{sec:4airtest}
\subsection{Hardware Setup}

A two-board IV-MUX (30 channels) was assembled and mounted inside an aluminum enclosure, as shown in Figure \ref{fig:airmux}. The IV-MUX was used in two configurations: in an IV configuration (mode B), where the current from the SiPM anode was measured, and in a pulse-counting configuration, where the IV-MUX was used to activate the bias voltage for the SiPMs and the output was measured dynamically. In pulse-counting mode, individual Geiger discharges are directly digitized. While current resolution per se is not a limiting factor, parasitic capacitance and poor grounding can degrade the pulse amplitude or introduce pickup. This is addressed through our shielding/guarding scheme, and through the separation of high-side (bias/pulse-counting) and low-side (current-sensing) paths. The purpose of this dual-configuration setup was to demonstrate that the multiplexed IV response is comparable to dynamic measurements. To enable both IV and pulse-counting measurements, the bias node of the IV-MUX was connected to a bias filter, which was wired in parallel with an SiPM and an amplifier. 

\begin{figure*}[t]
    \centering
    \includegraphics[width=\textwidth]{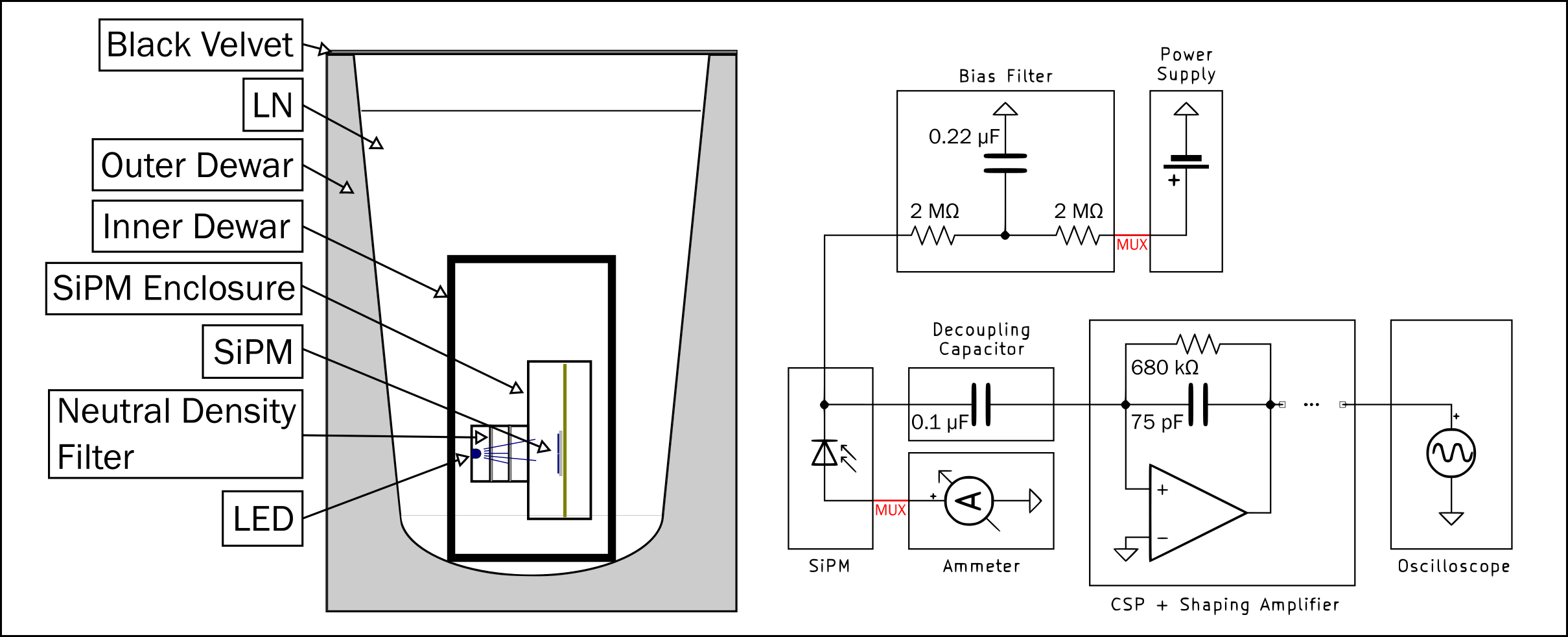}
    \caption[A schematic of the experimental setup and a simplified electrical diagram.]{Left: A schematic of the experimental setup. The SiPM is inside of an aluminum enclosure that is outfitted with an LED and neutral density filters. The enclosure is placed within two nested dewars and submerged in LN. Right: A simplified electrical diagram of a single channel for IV and pulse-counting measurements. The high-side of the SiPM is used for pulse-counting measurements, and the low-side for IV measurements. The bias node of the IV-MUX is connected to the bias filter, which is wired in parallel to the SiPM cathode and the decoupling capacitor (which is connected to the amplifier). The current sense node of the IV-MUX is connected to the anode of the SiPM.
    }
    \label{fig:muxexpschematic}
\end{figure*}

The electrometer (Keysight, B2987A) was used to measure current in the IV-configuration. The pulse-counting configuration used a charge-sensitive preamplifier and a 1 $\mathrm{\mu s}$ shaping amplifier (Cremat, CSP-112 and CR-200) with its output recorded using a digital oscilloscope (Rhode and Schwarz, RTO2024). IV and pulse-counting measurements of a 4-channel Hamamatsu S13371-6050CQ-02 SiPM \cite{ref:sipmvuv4} were recorded at 77 K, while submerged in LN. The SiPM was affixed to a PCB designed for compatibility with modes A and B of the multiplexer, similar to the array used in Section \ref{sec:3vactest}; however, the experiment reported in this section exclusively used the mode B configuration. The PCB was fixed within a small aluminum enclosure that had feedthroughs for four bias channels, the current-sensing line, and an optical port. A blue LED and two neutral density filters (Thorlabs, NDX0A) were mounted to the optical port to illuminate the SiPM with dim light pulses, enabling measurement of its electrical properties at faster rates than those achievable in dark conditions. The enclosure and SiPM is shown on the right in Figure \ref{fig:airmux}. The aluminum enclosure was placed in two nested dewars, and then covered with a lid and a black velvet blanket. A schematic of the experimental setup, and a simplified electrical diagram are shown in Figure \ref{fig:muxexpschematic}.

\subsection{Measurements}\label{sec:4airtest_meas}

\begin{figure}[t]
    \centering
    \includegraphics[width=\linewidth]{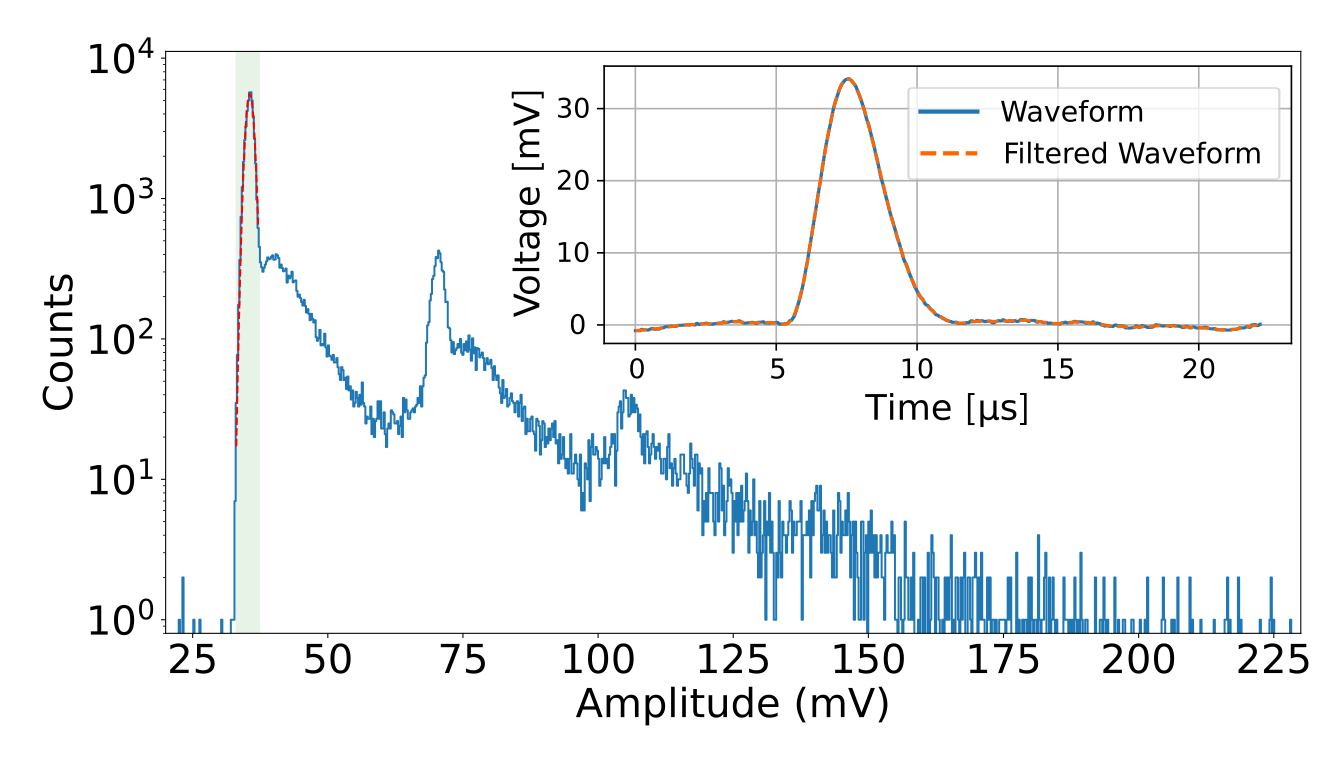}
    \caption{The uncalibrated charge spectrum for channel 2 of the SiPM at 5 V overvoltage. The SPE peak is centered around 35 mV and is fitted using a Gaussian function with asymmetric range constraints, as indicated by the green band. Peaks at integer multiples of 35 mV indicate crosstalk, while counts between these peaks result from afterpulsing. An example waveform of a typical SPE pulse is shown in the inset.}
    \label{fig:qspec}
\end{figure}

A simple relation is used to compare the IV measurements to the dynamic response \cite{ref:piemonte2012development}:

\begin{equation}\label{eq:dark_current}
    I_D = \mathrm{DCR} \cdot \mathrm{ECF} \cdot Q_0 + L,
\end{equation}

\noindent where $I_D$ is the dark current, $\mathrm{DCR}$ is the dark count rate, $\mathrm{ECF}$ is the excess-charge factor, $Q_0$ is the SPE charge, and $L$ is the leakage current. The ECF is the transfer function of an SiPM that converts input avalanches into output avalanches—in this case, the DCR—to account for all the extra charge from CAs. Technically, using the ECF to describe the current from an SiPM is not entirely appropriate because it diverges when afterpulses occur recursively, resulting in a finite DC current. An excess current factor is more appropriate; however, before reaching this critical point, where the ECF diverges, the ECF and the excess current factor are equal. The dark current and leakage current were measured using the electrometer, and the DCR, ECF, and SPE charge were measured using the oscilloscope. The DCR, ECF, and SPE charge were measured at nine bias voltages from 44 V to 52 V, and the dark current was measured from 38 V to 52 V in 0.1 V steps, with ten measurements at each step. The waveforms measured by the oscilloscope were digitized and analyzed using a custom Python script to determine the pulse amplitude and peak time. The charge spectrum was computed by extracting the pulse amplitude from a digitally filtered waveform and scaling the result by the amplifier gain, which was determined by applying a square wave to a calibrated capacitor to inject a known charge into the amplifier. The charge spectrum was analyzed to determine the SPE charge and ECF, and the timing distribution of the pulse peak time was analyzed to determine the DCR. Figure \ref{fig:qspec} shows the uncalibrated charge spectrum for SiPM 2 at a bias voltage of 47 V, with a typical SPE waveform displayed in the inset.

\begin{figure}[t]
    \centering
    \includegraphics[width=\columnwidth]{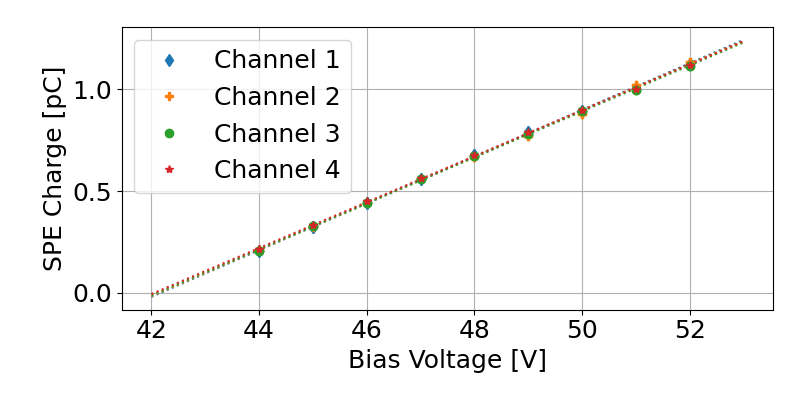}
    \caption{The voltage dependence of the SPE charge, measured for all channels of the SiPM. The error bars are, generally, smaller than the markers used in the plot. The y-error is calculated using the fit to the SPE peak position, and the x-error using the instrumental uncertainty.}
    \label{fig:gain}
\end{figure}

The SPE charge and ECF were measured with a blue LED to acquire a precise measurement in a shorter time, given the low DCR at cryogenic temperature ($\sim$ 1 cps). Both the SPE charge and the ECF were determined using the same data set of 100,000 waveforms, 22 $\mathrm{\mu}$s long. For a given bias voltage, the SPE charge was determined by fitting the SPE peak with a Gaussian function using asymmetric range constraints to prevent afterpulsing from introducing a bias to the fit. The voltage dependence of the SPE charge was modeled as a linear relationship:

\begin{equation}\label{eq:gain}
    Q_0 = C (V-V_{BD}),
\end{equation}

\noindent where $C$ is the effective SPAD capacitance, $V$ is the bias voltage, and $V_{BD}$ is the breakdown voltage. The breakdown voltage determined using Equation \ref{eq:gain} is used to calculate the overvoltage throughout the analysis, defined as the bias voltage exceeding the breakdown voltage. A plot of the SPE charge and the fit of Equation \ref{eq:gain} is shown in Figure \ref{fig:gain}.

The LED was driven with pulses shorter than the SiPM’s recovery time, which is around 50 ns, and the optics were configured such that a 50 kHz pulse rate resulted in an SiPM detection rate of nearly 100 cps with a 0.5 PE trigger threshold at 2 V overvoltage. The large separation between the likelihood of detection and non-detection indicates a negligible probability of detecting two photons from the LED. Therefore, any avalanches that occur within the acquisition window must be initiated by a single SPAD firing, similar to dark avalanches at cryogenic temperatures. This configuration guarantees that the ECF determined through this method is appropriate to apply to the DCR. Two methods were used to determine the ECF, which can only be considered equivalent if our assumption regarding a negligible contribution from two-photon detection is correct. Both methods rely on the following relation for the mean number of avalanches:

\begin{equation}
    \langle k \rangle = \mu \cdot \mathrm{ECF},
\end{equation}

\noindent where $\mu$ is the mean number of detections from the LED. The first method for determining the ECF assumes Poisson statistics to determine the mean number of detections by counting zero-detections through the following relation \cite{ref:otte2006measurement}:

\begin{equation}
    \mu = -\log{\left(1 - \frac{r}{f}\right)},
\end{equation}

\noindent where $r$ is the detection rate and $f$ is the pulse rate of the LED. The mean number of avalanches is determined by taking the mean of the charge spectrum, normalized by the SPE charge, and multiplied by the ratio of the detection rate to the flashing rate.

The second method for determining the ECF uses the assumption that the avalanches from the SiPM were initiated by a single SPAD firing, which means that $\mu$ is identically one for all events, and the mean number of avalanches is the ECF \cite{ref:jamil2018vuv}. The ECF derived using both methods was generally in agreement to better than 0.1\% accuracy, which validates our method for determining the equivalent ECF for the DCR.

\begin{figure}[t]
    \centering
    \includegraphics[width=\columnwidth]{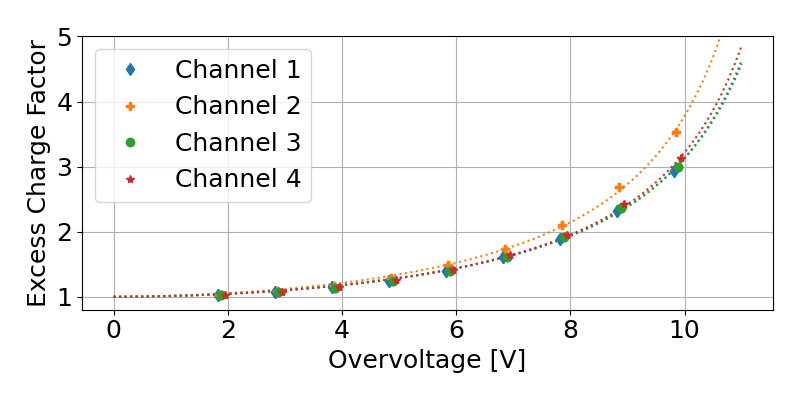}
    \caption{The voltage dependence of the ECF, measured for all channels of the SiPM. The ECF model has a singularity to account for the overvoltage where CAs occur recursively, known as the second divergence of an SiPM. The error bars are, generally, smaller than the markers used in the plot. The y-error is calculated using the combined error for both the number of avalanches and the number of detections, and the dominant contribution to the x-error comes from the uncertainty in the breakdown voltage.}
    \label{fig:ecf}
\end{figure}

The ECF is the SiPM's transfer function, accounting for the addition of CAs to the true detected signal. If the probability for CAs is low, then the ECF is 

\begin{equation}
    \mathrm{ECF} = 1 + \langle \Lambda \rangle,
\end{equation}

\noindent where $\langle \Lambda \rangle$ is the average extra charge accompanying a detection, i.e., the weighted average of all CAs and their respective probabilities. It is possible to model distinct processes for the different types of correlated avalanches \cite{ref:otte2017characterization}; however, this poses a risk of over-fitting the data due to insufficient degrees of freedom. At high overvoltage, recursive CAs occur when the probability for a CA exceeds unity, resulting in a divergent output from an input avalanche, known as the second divergence. A semi-empirical relation is used where the average extra charge is approximated using the product of a gain dependent term and a breakdown dependent term, and added geometrically \cite{ref:vinogradov2012analytical}:

\begin{equation}\label{eq:ecf}
    \mathrm{ECF} = \frac{1}{1-aV_{OV}\left( 1 - \exp{(-b V_{OV}}) \right)},
\end{equation}

\noindent where $a$ and $b$ are fit parameters. A plot of the ECF and the fit of Equation \ref{eq:ecf} is shown in Figure \ref{fig:ecf}.

\begin{figure}[t]
    \centering
    \includegraphics[width=\columnwidth]{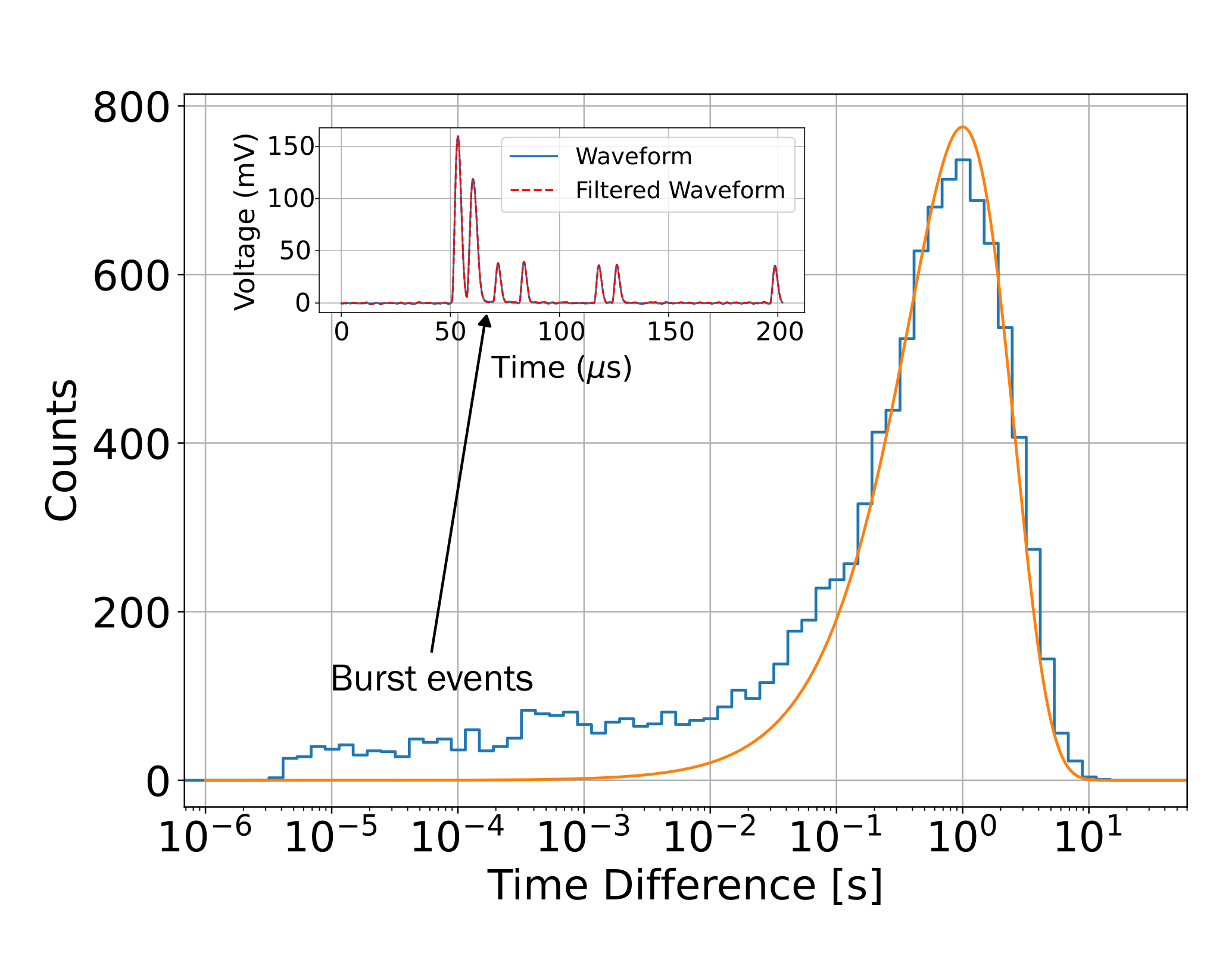}
    \caption{The time delay between all counts for channel 2 of the SiPM in the dark at 5 V overvoltage. An exponential distribution is fit to the histogram for counts with a time difference greater than 0.1 s. Excess counts with shorter time differences are due to burst events, likely triggered by a cosmic muon causing luminescence. The inset shows a waveform of a typical burst event.}
    \label{fig:spprob}
\end{figure}

The DCR is typically calculated by analyzing the interarrival times between uncorrelated avalanches. The time differences are binned in a histogram and an exponential distribution is fit to the data \cite{ref:piemonte2012development}. The DCR of the SiPMs used in this analysis are $\sim$ 1 cps at cryogenic temperature; however, burst events with multiple avalanches in a short time window were observed. Similar phenomena have been reported, attributed to cosmic muons or other ionizing particles causing luminescence \cite{ref:guarise2021newly, ref:tsang2023studies, ref:guarise2024investigation}. Our observations are comparable but involve significantly fewer events, likely due to the use of windowless devices, suggesting that the luminescence may originate from another material, such as the PCB, or a trap within the SiPM itself. We emphasize that the burst events are not afterpulses, as their timescale is much longer. The probability of detecting an afterpulse 1 $\mathrm{\mu s}$ after an avalanche is negligibly small for VUV4 devices at cryogenic temperatures \cite{ref:wang2021characterization, ref:pershing2022performance, ref:gallina2022performance}. Figure \ref{fig:spprob} shows the distribution of interarrival times and an exponential distribution fit to events with a time difference greater than 0.1 s; the waveform of a typical burst event is displayed in the inset. 

\begin{figure}[t]
    \centering
    \includegraphics[width=\columnwidth]{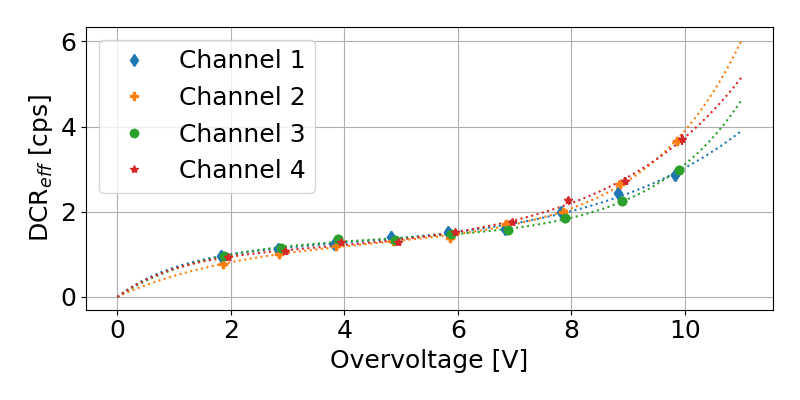}
    \caption{The voltage dependence of the DCR, measured for all channels of the SiPM. The error bars are, generally, smaller than the markers used in the plot. The y-error is calculated assuming Poisson fluctuations in the count rate; however, this underestimates the true error, as the avalanche rate in a burst event decays exponentially. The true measurement error is estimated using the best fit to Equation \ref{eq:dcr}, as described in Section \ref{sec:4airtest_meas}. The dominant contribution to the x-error comes from the uncertainty in the breakdown voltage.}
    \label{fig:dcr}
\end{figure}

\begin{figure*}[ht]
    \centering
    \includegraphics[width=\textwidth]{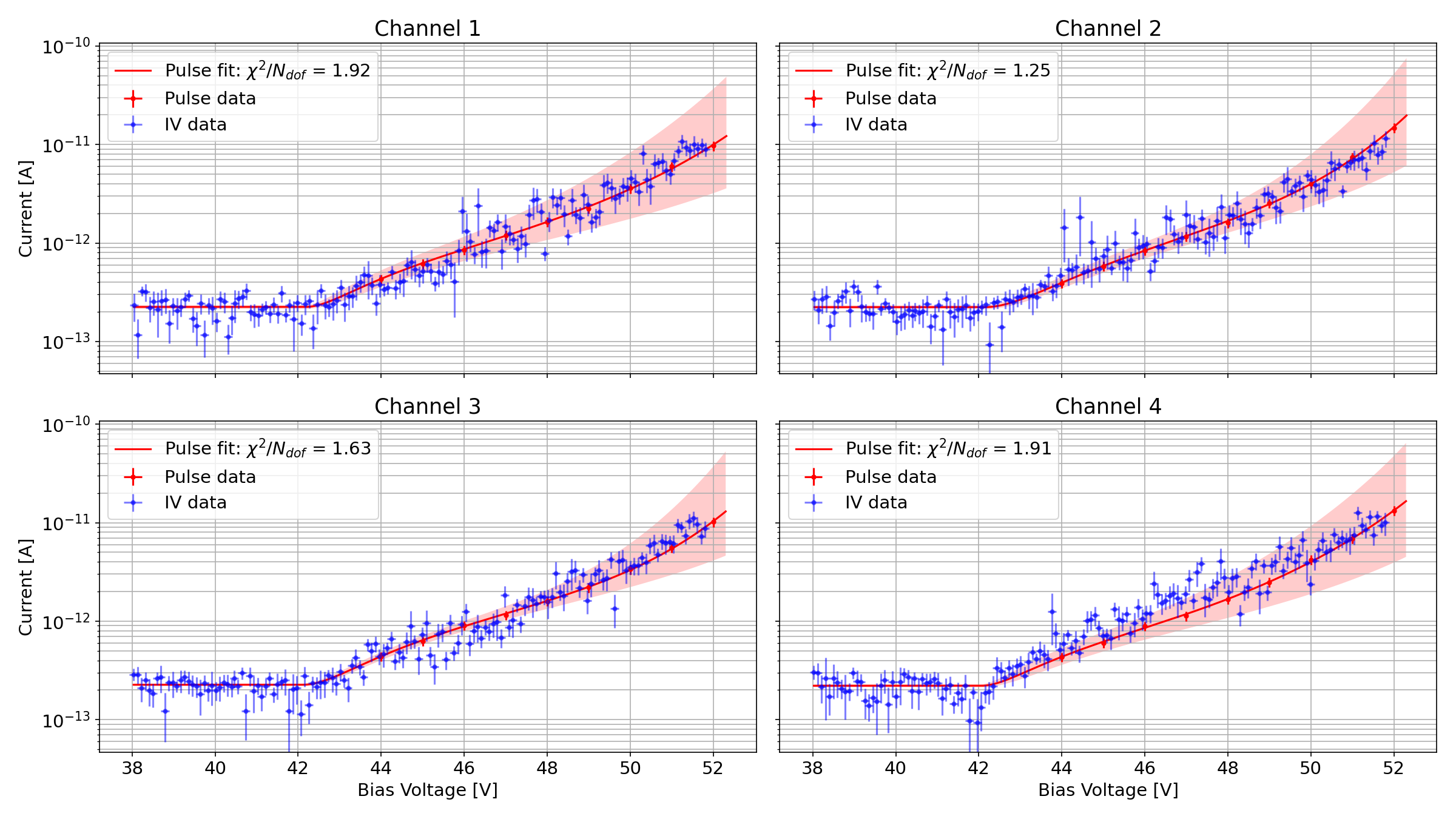}
    \caption{The dark current for each channel from the SiPM, as determined by both dynamic and static measurements. The current measured using the electrometer is shown with blue circles, the current calculated using Equation \ref{eq:dark_current} is shown with red circles, and the voltage dependent functional form of the dark current, determined through fits to the pulse-counting measurements, is shown with a red line. The red band represents the error in the functional form of the dark current, which increases significantly at high overvoltage due to uncertainty in predicting the second divergence. The average current in the pre-breakdown region was $224 \pm 9$ fA, $223 \pm 9$ fA, $226 \pm 6$ fA, and $221 \pm 7$ fA for channels 1-4, respectively. The maximum channel-to-channel variation in the average pre-breakdown current is $<$ 6 fA, which is consistent with no variation at our level of precision. The standard deviation in the pre-breakdown region was calculated for all four channels, and the average was $\sigma_{< \mathrm{BD}} = 49$ fA. }
    \label{fig:datamuxln2iv}
\end{figure*}

We must consider both the DCR and burst events, as they cannot be excluded from our IV measurements. A comparison between IV and pulse-counting requires an effective DCR, $\mathrm{DCR}_{eff}$, that includes all avalanches from the SiPM. The effective DCR was determined by acquiring 10,000 waveforms, each 200 $\mu$s long, recording the peak time for events above a 0.5 PE threshold, and then calculating the inverse of the mean time difference between all events. This method is not exact for two reasons. First, it includes late-arriving afterpulses in the determination of the effective DCR, even though their charge contribution is already accounted for in the ECF. Second, it assumes that all burst events are initiated by a single photon, which is not valid because the initial burst event rate is high enough for two or more photons to be detected within the amplifier's shaping time. However, these considerations introduce a sub-dominant error, as the probability of late-arriving afterpulses is low, and the total number of avalanches in a burst event is much greater than the likelihood-weighted contribution for two or more avalanches induced by the luminescence to occur within the amplifier's shaping time.

The voltage dependence of the effective DCR was initially modeled using the probability of thermally created electron-hole pairs, modulated by the avalanche triggering probability \cite{ref:otte2017characterization}. However, this approach failed to reproduce the observed trend due to the burst events. While burst events can, in principle, be tagged and modeled separately from dark events, doing so would introduce several new floating parameters, including the burst event rate, the number of detections per burst, and the spectral dependence on avalanche triggering probability. Since such modeling is beyond the scope of this paper, we use a simplified model with a single parameter to account for the mean detection rate of burst events, assuming the same avalanche triggering probability for all events,

\begin{equation}\label{eq:dcr}
    \mathrm{DCR}_{eff} = (R \exp{(\beta V_{OV})} + B) \left(1 - \exp{(-\alpha V_{OV})}\right),
\end{equation}

\noindent where $R$ is the dark count rate at breakdown, $\beta$ is the rate parameter, $B$ is the mean detection rate for burst events, $\alpha$ is the breakdown parameter, $V_{OV}$ is the overvoltage ($V - V_{BD}$). A plot showing the effective DCR and the fit of Equation \ref{eq:dcr} to the data is shown in Figure \ref{fig:dcr}. The error in the effective DCR is estimated by assigning Poisson fluctuations to each data point and then uniformly scaling each error until the reduced chi-square for the fit to Equation \ref{eq:dcr} equals one. This approach is used for its simplicity rather than calculating the errors for the DCR and burst events individually and combining them.

As shown in Equation \ref{eq:dark_current}, the SiPM dark current can be predicted using the DCR, SPE charge, and the ECF. We calculated the SiPM dark current using the aforementioned parameters, and computed a functional form of the dark current using the voltage dependent models. Figure \ref{fig:datamuxln2iv} shows the dynamically measured dark current, calculated using Equation \ref{eq:dark_current}, compared to the static dark current measured with the electrometer. The leakage current is calculated using the pre-breakdown region, and is added to the dynamically measured dark current as a constant offset. The average current in the pre-breakdown region was $224 \pm 9$ fA, $223 \pm 9$ fA, $226 \pm 6$ fA, and $221 \pm 7$ fA for channels 1-4, respectively. The standard deviation of the data in the pre-breakdown region is $\sigma_{< \mathrm{BD}}$ = 49 fA, which shows improvement compared to our tests inside the vacuum chamber. In general, the pulse-counting and IV methods show good agreement, demonstrating that integrating the IV-MUX does not degrade the expected small signal from an SiPM in the dark at cryogenic temperature. This validates both the utility of the device and the methods used herein for future measurements.

\section{Discussion}\label{sec:5discussion}
\subsection{Channel-to-Channel Variation and Calibration}

Since the IV-MUX switches each channel to a shared electrometer, all measurements are performed through the same low-noise readout chain. Channel-to-channel variation is primarily driven by relay leakage, cable capacitance, and PCB surface currents. These effects are reduced through guarded signal paths, separate bias and sense lines, and vacuum-compatible cleaning procedures to remove possible conductive surface contaminates. The maximum channel-to-channel variation in the average pre-breakdown current (Figure \ref{fig:datamuxln2iv}) is $<$ 6 fA, which is consistent with no variation at our level of precision. Thus, calibrating the absolute gain/offset for a single channel is sufficient for all channels. For current sensing devices, subtracting the un-stimulated signal from the stimulated signal removes any residual offset, e.g., subtracting the dark current from the current under illumination for a photodiode. To measure the dark current with Geiger mode devices, the pre-breakdown region provides a practical zero point for dark current measurements since the single-photoelectron gain increases sharply at breakdown. For devices that do not have a reference for subtraction, a dedicated channel tied to a calibration source can be used to monitor or correct for long-term drift and offset.

\subsection{Advantages, Limitations, and Novelty}

The IV-MUX enables automated high-sensitivity current measurements for up to 105 channels with a single ammeter, while maintaining sub-pA resolution. It achieves a minimum resolvable current of $<49\,\mathrm{fA}$ and supports both static IV and dynamic pulse-counting modes. The system is scalable to 105 channels per Arduino, compatible with vacuum and cryogenic environments, and is based on a low-cost, open-source design. Scaling up to higher channel numbers is possible with additional Arduinos or a different microcontroller. The primary limitations of the present implementation are that relay heating in vacuum approaches the maximum operating temperature, dynamic mode cannot be multiplexed, and the DPST relay footprint limits the achievable channel density on a $10\times10\,\mathrm{cm}^2$ PCB. More careful layout, following best practices for RF circuit design, could allow dynamic signals to be multiplexed into the microwave regime, and replacing the mechanical relays with solid-state relays would allow a significant reduction in board size.

\subsection{Performance Summary}

The main performance parameters of the IV-MUX, as demonstrated in the measurements presented here, are summarized in Table \ref{tab:specs}. These values represent the performance achieved during cryogenic and vacuum operation as demonstrated in the measurements presented in this work.

\begin{table}[H]
\centering
\caption{Summary of IV-MUX performance specifications}
\label{tab:specs}
\begin{tabular}{ll}
\hline
\textbf{Parameter} & \textbf{Value} \\
\hline
Max Channels                     & 105 per Arduino \\
Min. Resolvable Current          & $<49\,\mathrm{fA}$ \\
Channel-to-Channel Variation     & $<6\,\mathrm{fA}$ (pre-breakdown) \\
Bias Range                       & $\pm 200\,\mathrm{V}$ (relay dependent) \\
Relay Type                       & DPST + HI12 (high isolation) \\
Max Temperature (Vacuum)         & 350\,K \\
\hline
\end{tabular}
\end{table}

\subsection{Applicability}

The IV-MUX is suited for automated, low-current characterization and diagnostics of devices with many independent coaxial or triaxial connections accessible outside a vacuum or cryogenic environment. Typical use cases include large arrays of SiPMs, photodiodes, and radiation detector modules, where individual channels are wired through low-noise feedthroughs. The IV-MUX may serve as a diagnostic or commissioning tool in experiments with large detector arrays. During installation, cooldown, or scheduled maintenance periods—when the main DAQ is not operating or when rapid testing is preferred—the IV-MUX offers an efficient means of assessing functionality. Compared to dynamic pulse-counting, IV measurements are simpler to automate, faster to acquire, and require less data bandwidth and processing. This makes the system applicable for characterizing large detector arrays during commissioning, where rapid IV measurements reduce data volume and simplify analysis compared to dynamic readout. While not intended for continuous monitoring, the IV-MUX can supplement DAQ systems when scalable, standalone checks are required.

\section{Conclusion}\label{sec:6conclusion}

We presented the design and performance studies of a multiplexing system tailored for automated SiPM IV measurement, and showed that we could achieve a resolution as low as 49 fA when using this tool, with a channel-to-channel variation of no more than 6 fA. The IV-MUX can also be combined with a multichannel amplifier to enable both static and dynamic characterization of SiPMs, making it a versatile tool for comprehensive device evaluation across multiple operating modes. The IV-MUX is capable of routing the signals to a single ammeter, or as parallelized output to a multichannel ammeter, significantly reducing the cost and time required to measure IV curves of large SiPM arrays. The IV-MUX is vacuum compatible, reducing the cost per channel by eliminating the need for specialized multi-channel vacuum feedthroughs. Our results demonstrate that measuring the dark current with the IV-MUX is comparable to the prediction through dynamic measurements, even at cryogenic temperature (77 K) where the dark current is less than 1 pA. Integrating the IV-MUX into a measurement system can significantly reduce the measurement time through automated multiplexing, while still maintaining the accuracy of the IV measurement. Although the IV-MUX was designed for SiPM testing, it can be used for similar current-sensing devices where high channel count and low leakage current are required.

\section*{Acknowledgment}
This work was supported by the Natural Sciences and Engineering Research Council of Canada, the Canada Foundation for Innovation, and The CFREF - Arthur B. McDonald Institute. The authors would like to acknowledge support from the nEXO collaboration for supplying the SiPMs used for vacuum testing and the instrumentation division at Brookhaven National Lab for designing the SiPM carrier PCB, Fabrice Reti\`{e}re for discussions regarding SiPM IV reconstruction, and Serge Charlebois and Fr\'{e}d\'{e}ric Vachon for discussions regarding activation of defects/impurities in SiPM SPADs.

\bibliographystyle{IEEEtran} 
\bibliography{references}

\begin{thebibliography}{10}
\providecommand{\url}[1]{#1}
\csname url@samestyle\endcsname
\providecommand{\newblock}{\relax}
\providecommand{\bibinfo}[2]{#2}
\providecommand{\BIBentrySTDinterwordspacing}{\spaceskip=0pt\relax}
\providecommand{\BIBentryALTinterwordstretchfactor}{4}
\providecommand{\BIBentryALTinterwordspacing}{\spaceskip=\fontdimen2\font plus
\BIBentryALTinterwordstretchfactor\fontdimen3\font minus \fontdimen4\font\relax}
\providecommand{\BIBforeignlanguage}[2]{{%
\expandafter\ifx\csname l@#1\endcsname\relax
\typeout{** WARNING: IEEEtran.bst: No hyphenation pattern has been}%
\typeout{** loaded for the language `#1'. Using the pattern for}%
\typeout{** the default language instead.}%
\else
\language=\csname l@#1\endcsname
\fi
#2}}
\providecommand{\BIBdecl}{\relax}
\BIBdecl

\bibitem{ref:renker2009advances}
D.~Renker and E.~Lorenz, ``Advances in solid state photon detectors,'' \emph{Journal of Instrumentation}, vol.~4, no.~04, p. P04004, 2009.

\bibitem{ref:ghassemi2017mppc}
A.~Ghassemi, K.~Sato, and K.~Kobayashi, ``{MPPC},'' 2017, \url{https://www.hamamatsu.com/content/dam/hamamatsu-photonics/sites/documents/99_SALES_LIBRARY/ssd/mppc_kapd9005e.pdf}.

\bibitem{ref:acerbi2019understanding}
F.~Acerbi and S.~Gundacker, ``{Understanding and simulating SiPMs},'' \emph{Nuclear Instruments and Methods in Physics Research Section A: Accelerators, Spectrometers, Detectors and Associated Equipment}, vol. 926, pp. 16--35, 2019.

\bibitem{ref:kharusi2018nexo}
S.~Al~Kharusi, A.~Alamre, J.~Albert, M.~Alfaris, G.~Anton, I.~Arnquist, I.~Badhrees, P.~Barbeau, D.~Beck, V.~Belov \emph{et~al.}, ``{nEXO pre-conceptual design report},'' \emph{{arXiv preprint arXiv:1805.11142}}, 2018.

\bibitem{ref:aalseth2018darkside}
C.~E. Aalseth, F.~Acerbi, P.~Agnes, I.~Albuquerque, T.~Alexander, A.~Alici, A.~Alton, P.~Antonioli, S.~Arcelli, R.~Ardito \emph{et~al.}, ``{DarkSide-20k: A 20 tonne two-phase LAr TPC for direct dark matter detection at LNGS},'' \emph{The European Physical Journal Plus}, vol. 133, pp. 1--129, 2018.

\bibitem{ref:dune2020deep}
B.~Abi \emph{et~al.}, ``{Deep Underground Neutrino Experiment (DUNE): Far detector technical design report. Volume I. Introduction to DUNE},'' \emph{Journal of Instrumentation}, vol.~15, no.~8, 2020.

\bibitem{ref:otte2017characterization}
A.~Otte, D.~Garcia, T.~Nguyen, and D.~Purushotham, ``Characterization of three high efficiency and blue sensitive silicon photomultipliers,'' \emph{Nuclear Instruments and Methods in Physics Research Section A: Accelerators, Spectrometers, Detectors and Associated Equipment}, vol. 846, pp. 106--125, 2017.

\bibitem{gallina2019characterization}
G.~Gallina, F.~Reti{\`e}re, P.~Giampa, J.~Kroeger, P.~Margetak, S.~B. Mamahit, A.~D.~S. Croix, F.~Edaltafar, L.~Martin, N.~Massacret \emph{et~al.}, ``Characterization of {SiPM} avalanche triggering probabilities,'' \emph{{IEEE Transactions on Electron Devices}}, vol.~66, no.~10, pp. 4228--4234, 2019.

\bibitem{gallina2021development}
G.~Gallina, ``Development of a single vacuum ultra-violet photon-sensing solution for {nEXO},'' Ph.D. dissertation, University of British Columbia, 2021.

\bibitem{ref:gallina2019characterization}
G.~Gallina, P.~Giampa, F.~Reti{\`e}re, J.~Kroeger, G.~Zhang, M.~Ward, P.~Margetak, G.~Li, T.~Tsang, L.~Doria \emph{et~al.}, ``Characterization of the {Hamamatsu} {VUV4} {MPPCs} for {nEXO},'' \emph{Nuclear Instruments and Methods in Physics Research Section A: Accelerators, Spectrometers, Detectors and Associated Equipment}, vol. 940, pp. 371--379, 2019.

\bibitem{ref:wang2021characterization}
L.~Wang, M.~Guan, H.~Qin, C.~Guo, X.~Sun, C.~Yang, Q.~Zhao, J.~Liu, P.~Zhang, Y.~Zhang \emph{et~al.}, ``Characterization of {VUV4} {SiPM} for liquid argon detector,'' \emph{Journal of instrumentation}, vol.~16, no.~07, p. P07021, 2021.

\bibitem{ref:alvarez2024measurement}
R.~{\'A}lvarez-Garrote, E.~Calvo, A.~Canto, J.~I. Crespo-Anad{\'o}n, C.~Cuesta, A.~de~la Torre~Rojo, I.~Gil-Botella, S.~M. Corchado, I.~Mart{\'\i}n, C.~Palomares \emph{et~al.}, ``Measurement of the photon detection efficiency of {Hamamatsu} {VUV4} {SiPMs} at cryogenic temperature,'' \emph{Nuclear Instruments and Methods in Physics Research Section A: Accelerators, Spectrometers, Detectors and Associated Equipment}, vol. 1064, p. 169347, 2024.

\bibitem{ref:borden2024characterization}
S.~Borden, J.~Detwiler, W.~Pettus, and N.~Ruof, ``Characterization of silicon photomultiplier photon detection efficiency at liquid nitrogen temperature,'' \emph{arXiv preprint arXiv:2405.01529}, 2024.

\bibitem{ref:piemonte2019overview}
C.~Piemonte and A.~Gola, ``Overview on the main parameters and technology of modern {Silicon Photomultipliers},'' \emph{Nuclear Instruments and Methods in Physics Research Section A: Accelerators, Spectrometers, Detectors and Associated Equipment}, vol. 926, pp. 2--15, 2019.

\bibitem{cova1996avalanche}
S.~Cova, M.~Ghioni, A.~Lacaita, C.~Samori, and F.~Zappa, ``Avalanche photodiodes and quenching circuits for single-photon detection,'' \emph{Applied optics}, vol.~35, no.~12, pp. 1956--1976, 1996.

\bibitem{giudice2007high}
A.~Giudice, M.~Ghioni, R.~Biasi, F.~Zappa, S.~Cova, P.~Maccagnani, and A.~Gulinatti, ``High-rate photon counting and picosecond timing with silicon-{SPAD} based compact detector modules,'' \emph{Journal of Modern Optics}, vol.~54, no. 2-3, pp. 225--237, 2007.

\bibitem{parent2024wafer}
S.~Parent, F.~Vachon, V.~Gauthier, S.~Lamoureux, A.~Paquette, J.~Deschamps, T.~Rossignol, N.~Roy, P.~Arsenault, H.~Dautet \emph{et~al.}, ``Wafer-level characterization and monitoring platform for single-photon avalanche diodes,'' \emph{{IEEE Journal of the Electron Devices Society}}, 2024.

\bibitem{engelmann2018spatially}
E.~Engelmann, E.~Popova, and S.~Vinogradov, ``Spatially resolved dark count rate of {SiPMs},'' \emph{The European Physical Journal C}, vol.~78, pp. 1--8, 2018.

\bibitem{mclaughlin2021spatially}
J.~B. McLaughlin, G.~Gallina, F.~Reti{\`e}re, A.~De~St.~Croix, P.~Giampa, M.~Mahtab, P.~Margetak, L.~Martin, N.~Massacret, J.~Monroe \emph{et~al.}, ``Characterisation of {SiPM} photon emission in the dark,'' \emph{Sensors}, vol.~21, no.~17, p. 5947, 2021.

\bibitem{ref:acerbi2023nuv}
F.~Acerbi, G.~Paternoster, S.~Merzi, N.~Zorzi, and A.~Gola, ``{NUV} and {VUV} sensitive {Silicon} {Photomultipliers} technologies optimized for operation at cryogenic temperatures,'' \emph{Nuclear Instruments and Methods in Physics Research Section A: Accelerators, Spectrometers, Detectors and Associated Equipment}, vol. 1046, p. 167683, 2023.

\bibitem{ref:kester1992amplifierappguide}
W.~Kester, Ed., \emph{{Amplifier Applications Guide}}.\hskip 1em plus 0.5em minus 0.4em\relax Analog Devices, 1992.

\bibitem{ref:muxgit}
E.~Eegan and L.~Darroch, ``{IV-MUX-public},'' \url{https://github.com/Brunner-neutrino-lab/IV-MUX-public}, 2024.

\bibitem{ref:campbell2004recent}
J.~C. Campbell, S.~Demiguel, F.~Ma, A.~Beck, X.~Guo, S.~Wang, X.~Zheng, X.~Li, J.~D. Beck, M.~A. Kinch \emph{et~al.}, ``Recent advances in avalanche photodiodes,'' \emph{IEEE Journal of selected topics in quantum electronics}, vol.~10, no.~4, pp. 777--787, 2004.

\bibitem{ref:sipmvuv4}
\BIBentryALTinterwordspacing
\emph{{MPPC for MEGII experiment}}, Hamamatsu, 2017. [Online]. Available: \url{https://hamamatsu.su/files/uploads/pdf/3_mppc/s13370_vuv4-mppc_b_(1).pdf}
\BIBentrySTDinterwordspacing

\bibitem{ref:piemonte2012development}
C.~Piemonte, A.~Ferri, A.~Gola, A.~Picciotto, T.~Pro, N.~Serra, A.~Tarolli, and N.~Zorzi, ``Development of an automatic procedure for the characterization of silicon photomultipliers,'' in \emph{{2012 IEEE Nuclear Science Symposium and Medical Imaging Conference Record (NSS/MIC)}}.\hskip 1em plus 0.5em minus 0.4em\relax IEEE, 2012, pp. 428--432.

\bibitem{ref:otte2006measurement}
A.~Otte, J.~Hose, R.~Mirzoyan, A.~Romaszkiewicz, M.~Teshima, and A.~Thea, ``A measurement of the photon detection efficiency of silicon photomultipliers,'' \emph{Nuclear Instruments and Methods in Physics Research Section A: Accelerators, Spectrometers, Detectors and Associated Equipment}, vol. 567, no.~1, pp. 360--363, 2006.

\bibitem{ref:jamil2018vuv}
A.~Jamil, T.~Ziegler, P.~Hufschmidt, G.~Li, L.~Lupin-Jimenez, T.~Michel, I.~Ostrovskiy, F.~Reti{\`e}re, J.~Schneider, M.~Wagenpfeil \emph{et~al.}, ``{VUV-sensitive silicon photomultipliers for xenon scintillation light detection in nEXO},'' \emph{IEEE Transactions on Nuclear Science}, vol.~65, no.~11, pp. 2823--2833, 2018.

\bibitem{ref:vinogradov2012analytical}
S.~Vinogradov, ``Analytical models of probability distribution and excess noise factor of solid state photomultiplier signals with crosstalk,'' \emph{Nuclear Instruments and Methods in Physics Research Section A: Accelerators, Spectrometers, Detectors and Associated Equipment}, vol. 695, pp. 247--251, 2012.

\bibitem{ref:guarise2021newly}
M.~Guarise, M.~Andreotti, R.~Calabrese, A.~C. Ramusino, V.~Cicero, M.~Fiorini, T.~Giammaria, I.~Lax, E.~Luppi, A.~Minotti \emph{et~al.}, ``A newly observed phenomenon in the characterisation of {SiPM} at cryogenic temperature,'' \emph{Journal of Instrumentation}, vol.~16, no.~10, p. T10006, 2021.

\bibitem{ref:tsang2023studies}
T.~Tsang, H.~Chen, S.~Gao, G.~Giacomini, V.~Radeka, and S.~Rescia, ``{Studies of event burst phenomenon with SiPMs in liquid nitrogen},'' \emph{Journal of Instrumentation}, vol.~18, no.~01, p. C01050, 2023.

\bibitem{ref:guarise2024investigation}
M.~Guarise, M.~Andreotti, A.~Balboni, R.~Calabrese, D.~Casazza, A.~C. Ramusino, A.~Corallo, S.~Chiozzi, R.~D'Amico, M.~Fiorini \emph{et~al.}, ``Investigation of the burst phenomenon in {SiPMs} at liquid nitrogen temperature,'' \emph{arXiv preprint arXiv:2405.15922}, 2024.

\bibitem{ref:pershing2022performance}
T.~Pershing, J.~Xu, E.~Bernard, J.~Kingston, E.~Mizrachi, J.~Brodsky, A.~Razeto, P.~Kachru, A.~Bernstein, E.~Pantic \emph{et~al.}, ``Performance of hamamatsu vuv4 sipms for detecting liquid argon scintillation,'' \emph{Journal of Instrumentation}, vol.~17, no.~04, p. P04017, 2022.

\bibitem{ref:gallina2022performance}
G.~Gallina, Y.~Guan, F.~Reti{\`e}re, G.~Cao, A.~Bolotnikov, I.~Kotov, S.~Rescia, A.~K. Soma, T.~Tsang, L.~Darroch \emph{et~al.}, ``Performance of novel vuv-sensitive silicon photo-multipliers for nexo,'' \emph{The European Physical Journal C}, vol.~82, no.~12, p. 1125, 2022.

\end{thebibliography}

\vfill

\end{document}